\documentclass[aps, prc, reprint, amsmath, amssymb, nofootinbib, superscriptaddress, floatfix]{revtex4-2}
\usepackage{physics}
\usepackage{bbm}
\usepackage{bm}
\usepackage{graphicx}
\usepackage[table,xcdraw]{xcolor}
\usepackage[caption=false]{subfig}
\usepackage{ulem}
\usepackage{booktabs}
\usepackage{multirow}
\usepackage{url}
\usepackage[colorlinks=true, citecolor=blue, urlcolor=blue, linkcolor=blue]{hyperref}
\usepackage{natbib}
\newcommand{\Msun}{M_\odot}
\newcommand{\FD}{F_{\rm D}}
\newcommand{\mx}{m_\chi}
\newcommand{\cad}{c_{\rm s}}
\newcommand{\ceq}{c_{\rm eq}}

\newcommand{\bMc}{1.1872}

\newcommand{\qpOnA}{0.026}

\newcommand{\qpOnC}{0.17}

\newcommand{\qplnBmin}{-1.0}
\newcommand{\qplnBmax}{-0.6}
\newcommand{\qpUbmin}{0.12}
\newcommand{\qpUbmax}{0.26}

\newcommand{\qpRlo}{11.6}
\newcommand{\qpRhi}{13.2}

\newcommand{\qpLlo}{298}
\newcommand{\qpLhi}{626}
\newcommand{\qpUbSkC}{0.12}
\newcommand{\qpWrmf}{75}

\newcommand{\qpHaloMax}{1.2}

\newcommand{\qfRelC}{10}
\newcommand{\qfRelE}{15}
\newcommand{\qfHzC}{2}
\newcommand{\qfHzE}{3}

\newcommand{\qfRise}{35}
\newcommand{\qfDirFrac}{35}
\newcommand{\qmDflNuc}{0.05}
\newcommand{\qmDflC}{1.3}
\newcommand{\qmDflE}{0.7}
\newcommand{\qmDflEmax}{0.5}
\newcommand{\qmTwN}{150}
\newcommand{\qmTwAbove}{75}
\newcommand{\qmTwMed}{1.3}
\newcommand{\qmFitPct}{13}
\newcommand{\qmFitRPct}{27}
\newcommand{\qmdLC}{9}
\newcommand{\qmdLE}{12}

\newcommand{\qmDfCmin}{10}
\newcommand{\qmDfCmax}{14}
\newcommand{\qmDgCmin}{5}
\newcommand{\qmDgCmax}{17}
\newcommand{\qmDDflCmin}{1.1}
\newcommand{\qmDDflCmax}{1.6}

\newcommand{\qmDgEmax}{30}
\newcommand{\qfcGD}{+0.3}
\newcommand{\qfcGN}{-0.5}
\newcommand{\qfcGFD}{+13}
\newcommand{\qfcGFN}{-2.4}
\newcommand{\qlooGfrac}{5}

\newcommand{\qlooGFfrac}{27}
\newcommand{\qlooGFmax}{196}
\newcommand{\qdphiNuc}{0.002}
\newcommand{\qdphiCore}{0.006}
\newcommand{\qdphiNuchi}{0.03}
\newcommand{\qdphiCorehi}{0.03}
\newcommand{\qTauLo}{7}
\newcommand{\qTauHi}{12}

\begin{document}
\title{Composition $g$-modes and $f$-modes of neutron stars with gravitationally coupled dark matter: degeneracy with the symmetry energy}
\author{Probit J Kalita}
\affiliation{Department of Physics \& Astronomy, National Institute of Technology, Rourkela 769008, India}
\author{Bharat Kumar}
\email{kumarbh@nitrkl.ac.in}
\affiliation{Department of Physics \& Astronomy, National Institute of Technology, Rourkela 769008, India}
\author{H. C. Das}
\affiliation{$^2$Institut f\"ur Theoretische Physik, Goethe Universit\"at, Max-von-Laue-Str.~1, D-60438 Frankfurt am Main, Germany}

\date{\today}

\begin{abstract}
	Composition gravity ($g$) modes have been proposed as a probe of dark matter in neutron stars, on the basis of one-fluid models in which the dark matter is in chemical equilibrium with the neutrons. We computed in full general relativity the $g$- and $f$-modes of stars in which fermionic dark matter couples to the nucleons only through gravity, and inferred the dark mass fraction from NICER and GW170817 data with a prior that comprises $63$ nucleonic equations of state (EOSs) from Skyrme and relativistic mean-field functionals, among them the models of Dutra \textit{et al.} and of the compilation of Sun, Bhattiprolu and Lattimer. The data do not prefer dark matter and bound its mass fraction at $\qpUbmin$--$\qpUbmax$ ($95\%$) for particle masses of $0.5$--$2$~GeV. Coupled only through gravity, the dark matter reaches the buoyancy that restores a $g$-mode only through the gravitational field: a dark core holding $10\%$ of the mass raises the fundamental $g$-mode by $\qfRelC$--$\qfRelE\%$ at fixed EOS, several times less than in the one-fluid picture. The larger effect is indirect: a star with a dark core needs a stiffer nucleonic EOS, and since the $g$-mode is set by the density dependence of the symmetry energy, its frequency reflects the EOS that the star requires rather than the dark matter it contains; at fixed radius and symmetry energy a core holding $10\%$ of the mass is equivalent to a change of the slope $L$ of $\qmdLC$--$\qmdLE$~MeV. The $f$-mode, which follows the mean density, behaves differently: a dark core breaks the relation between the $f$-mode and the tidal deformability by about one percent, far outside the posterior spread of nucleonic stars.
\end{abstract}

\maketitle

\section{Introduction}

Neutron stars can carry dark matter captured from the Galactic halo or produced inside them~\cite{ellis2018,nelson2019,kain2021}. If the dark matter interacts with ordinary matter only through gravity, the star consists of two fluids with separately conserved stress tensors~\cite{leung2011,kain2021}, and the dark matter settles in a core or in an extended halo, depending on its particle mass, self-interaction and amount~\cite{ivanytskyi2020,miao2022,giangrandi2023}. The mass, radius and tidal deformability do not separate such a star from a nucleonic one: a stiffer nucleonic equation of state (EOS) with a dark core reproduces a softer one without~\cite{rutherford2023,thakur2024,rutherford2025}, and probes beyond the bulk properties are needed~\cite{rutherford2026}.

Composition gravity ($g$) modes respond to the interior composition rather than to the stiffness. Their restoring force is the buoyancy of a fluid element whose proton and lepton fractions stay frozen on the oscillation time scale~\cite{reisenegger1992}; their frequencies, $\sim100$--$500$~Hz, are set by the difference between the frozen-composition sound speed $\cad$ and the equilibrium sound speed $\ceq$, and hence by the density dependence of the symmetry energy~\cite{jaikumar2021,zhao2022,sotani2026}. The connection is direct: $\beta$ equilibrium fixes the proton fraction through $\mu_n-\mu_p\simeq4E_{\rm sym}(n)(1-2Y_p)$, so a symmetry energy that rises steeply with density, a large slope $L$, makes the proton fraction climb steeply toward the centre. A fluid element displaced downward then carries too few protons for its new surroundings, is lighter than they are at the same pressure, and is pushed back~\cite{reisenegger1992}; the steeper the proton-fraction gradient, the stronger this buoyancy and the higher the $g$-mode frequency. They can be excited resonantly during a binary inspiral~\cite{lai1994,ho1999,yu2017,counsell2024}. For dark matter produced by neutron decay, which is in chemical equilibrium with the neutrons and forms one fluid with them, Shirke, Chatterjee and Jaikumar~\cite{shirke2025} found in the Cowling approximation that $g_1$ rises by $14.23$~Hz per percent of dark matter, nearly independently of the nucleonic EOS, and noted that a stiffer nucleonic EOS with dark matter can give the same frequency as a nucleonic star. For dark matter coupled only by gravity, the $f$-modes have been computed in the two-fluid Cowling approximation~\cite{thakur2024jcap,sotani2025a,sotani2025b}, where the $f$--tidal-deformability relation breaks for dark fractions of $0.2$--$0.6$~\cite{sotani2025b}, and in full general relativity (GR)~\cite{kumar2026,zheng2026}. The $g$-modes of such stars have not been computed.

In this paper we compute the $g$- and $f$-modes of two-fluid stars in full GR, infer the dark fraction from NICER and GW170817 data, and propagate the posterior to the mode frequencies. The picture is simple. Gravity alone couples the dark matter only weakly to the buoyancy, so the direct effect on the $g$-modes is small; the larger effect is indirect, through the stiffer nucleonic EOS that the data demand once the star carries dark matter, and it is therefore degenerate with the symmetry energy. The $f$-mode, which follows the mean density, provides a test that the nuclear model does not mimic.

\section{Model and method}

\subsection{Two-fluid stars}

Each fluid $i\in\{{\rm N},{\rm D}\}$ obeys $dp_i/dr=-(\varepsilon_i+p_i)\Phi'$ in the common metric $ds^2=-e^{2\Phi}dt^2+e^{2\Lambda}dr^2+r^2d\Omega^2$~\cite{leung2011,kain2021,zheng2026}, and $\FD=M_{\rm D}/M$ is the dark fraction of the gravitational mass. The nucleonic fluid is cold $npe\mu$ matter in $\beta$ equilibrium. We describe it by $63$ functionals, all with the same BSk24 crust~\cite{pearson2018}: $38$ Skyrme functionals from the compilation of Ref.~\cite{sun2024} (maximum mass $\ge2\,\Msun$, causal and stably stratified cores) and $25$ relativistic mean-field (RMF) functionals, namely six widely used models, NL3~\cite{lalazissis1997}, FSUGarnet~\cite{chen2015}, IOPB-I~\cite{kumar2018}, G3~\cite{kumar2017g3}, BigApple~\cite{fattoyev2020} and IUFSU~\cite{fattoyev2010}; the nine models with $\sigma$--$\omega$ and $\sigma$--$\rho$ cross couplings that Dutra \textit{et al.} found consistent with nuclear-matter constraints~\cite{dutra2014}, BKA20--24~\cite{agrawal2010}, BSR8--12~\cite{dhiman2007} and G2$^*$~\cite{sulaksono2006}; and ten models of the compilation of Ref.~\cite{sun2024} whose saturation properties lie in the same ranges and whose maximum mass exceeds $2\,\Msun$, BSR1--5~\cite{dhiman2007}, FSU2H~\cite{tolos2017}, PCSB1~\cite{pradhan2023npa}, QMC-RMF3~\cite{alford2022}, HC~\cite{bunta2003} and RMF412~\cite{dadi2010}, with the couplings tabulated there. The buoyancy of the nucleonic fluid is carried by
\begin{equation}
	\mathcal N^2=e^{2\Phi-2\Lambda}\Phi'^2D,\qquad D=\frac{1}{\ceq^2}-\frac{1}{\cad^2}\ge0,
	\label{eq:N2}
\end{equation}
with $\cad^2=(\partial p/\partial\varepsilon)_{Y_p,Y_e,Y_\mu}$~\cite{sotani2026}. The composition term $D$ belongs to the nucleons alone; the dark matter can enter $\mathcal N$ only through the gravitational field $\Phi'$. The dark fluid is a cold gas of fermions of mass $\mx=0.5$--$2$~GeV with a repulsive Yukawa self-interaction fixed by $\sigma_{\rm D}/\mx=1$~cm$^2$\,g$^{-1}$~\cite{zheng2026}, close to the Bullet Cluster limit~\cite{randall2008}. Whether the dark matter forms a core or a halo is decided by its Fermi pressure. A self-gravitating degenerate Fermi gas has a characteristic mass and radius that scale as $\mx^{-2}$, so at a given dark mass light particles are supported by a large Fermi pressure and spread beyond the nucleonic fluid, an extended halo ($R_{\rm D}>R_{\rm N}$), whereas heavy particles collapse into a compact core; the repulsive self-interaction adds pressure and moves the boundary to heavier particles. We solved the polar perturbation equations of the two gravitationally coupled fluids in the Lindblom--Detweiler variables~\cite{lindblom1983,detweiler1985,kumar2026,zheng2026}, with $1/c_{\rm s,N}^2=1/c_{\rm eq,N}^2-D$, matched to an outgoing Zerilli wave, and seeded the $g$-mode roots with the relativistic Cowling spectrum; every star discussed at $1.4\,\Msun$ (or $1.2\,\Msun$) was solved at exactly that mass. Our $f$- and $p_1$-modes reproduce the polytropic results of Ref.~\cite{dimmelmeier2006} and, for one fluid, the $f$--Love relation of Ref.~\cite{chan2014} to better than $0.5\%$. The $g$-modes radiate so weakly that we took their damping from the Zerilli luminosity and the mode energy~\cite{thorne1969a,zheng2026}.

\subsection{Bayesian analysis}

The model parameters are the nucleonic EOS $e$, one of the $63$ functionals with equal prior weight, and the dark fraction $\FD$, the same for all stars, with a flat prior on $[0,0.3]$; the likelihood was computed at ten values of $\FD$ and interpolated linearly between them, and each $\mx$ was analysed separately. The posterior is $p(e,\FD|d)\propto\mathcal L_{\rm GW}\prod_k\mathcal L_k$, with
\begin{align}
	\mathcal L_k        & =\int_{1\,\Msun}^{M_{\max}}\frac{\hat p_k[M,R_{\rm N}(M)]\,dM}{M_{\max}-1\,\Msun},\nonumber \\
	\mathcal L_{\rm GW} & =\int_{0.7}^{1}\frac{\hat p_{\rm GW}[q,\Lambda(m_1),\Lambda(m_2)]\,dq}{0.3}.
	\label{eq:like}
\end{align}
Here $\hat p_k$ is a Gaussian kernel density estimate (Scott bandwidth) of all equal-weight mass--radius samples of PSR J0030+0451 ($34\,120$, ST+PST)~\cite{riley2019}, J0437$-$4715 ($218\,553$, CST+PDT)~\cite{choudhury2024} and J0740+6620 ($379\,669$, NICER+XMM)~\cite{salmi2024}, evaluated at the radius of the nucleonic fluid, which carries the hot spots, with a flat mass prior up to the maximum mass of the model. $\hat p_{\rm GW}$ is a kernel density estimate of all $3952$ low-spin samples of $(q,\Lambda_1,\Lambda_2)$ of GW170817~\cite{abbott2019prx}, reflected at $q=1$ and $\Lambda_i=0$, with $m_{1,2}(q)$ fixed by the median source-frame chirp mass, $\bMc\,\Msun$. We used the published posteriors as likelihoods, which neglects the priors of the original analyses. The Bayes factor $\mathcal B$ is the ratio of the evidence with $\FD$ free to that with $\FD=0$. All results below are posterior quantities of this one prior.

\begin{figure}[t]
	\includegraphics[width=\columnwidth]{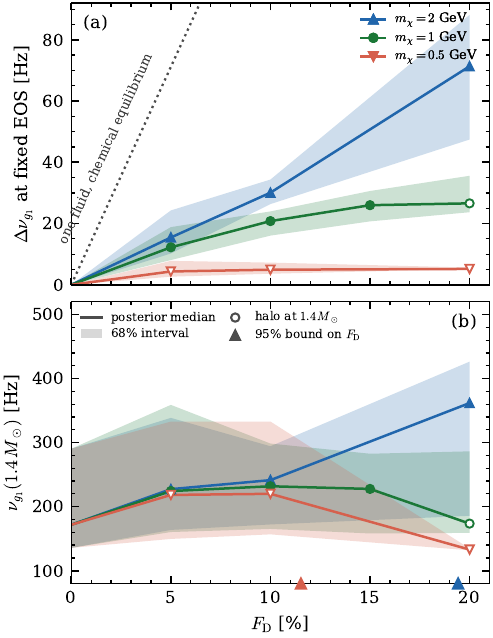}
	\caption{Fundamental $g$-mode of a $1.4\,\Msun$ star against the dark mass fraction, weighted at each $\FD$ by the NICER and GW170817 posterior over the nucleonic EOS (median and $68\%$ interval). (a) Shift at fixed EOS, relative to the nucleonic star of the same EOS; (b) the frequency itself. Filled symbols: the $1.4\,\Msun$ star has a dark core; open: a halo. Triangles: $95\%$ upper bounds on $\FD$. Dotted: $14.23$~Hz per percent, the one-fluid Cowling result for dark matter in chemical equilibrium with the neutrons~\cite{shirke2025}.}
	\label{fig:shift}
\end{figure}

\section{Results}

\subsection{A small direct shift and a larger indirect one}

Figure~\ref{fig:shift}(a) shows how much the dark matter itself moves $g_1$. At fixed EOS a dark core that holds ten percent of the mass raises the frequency by $\qfRelC$--$\qfRelE\%$, $\qfHzC$--$\qfHzE$~Hz per percent of dark matter at the posterior median, and up to $\qmDgEmax\%$ for the EOSs with the strongest buoyancy; a halo moves it by only a few hertz, whatever its mass. Both stay below the $14$~Hz per percent of the one-fluid picture~\cite{shirke2025}, and the reason is Eq.~(\ref{eq:N2}). A core deepens the potential in which the nucleons sit and raises $\Phi'$, and with it $\mathcal N$, but it leaves the composition gradient $D$ untouched; most of the mass of a halo lies outside the nucleonic fluid, where it does not enter $\Phi'$ at all. When the dark matter is in chemical equilibrium with the neutrons, by contrast, its abundance changes with depth and adds to the composition gradient itself~\cite{shirke2025}. The shift is largest for EOSs whose buoyancy is already strong (large $L$, high $g_1$) and smallest for soft ones such as SLy4.

Figure~\ref{fig:shift}(b) shows what the data do with this. Once the stars carry dark matter, NICER and GW170817 favour stiffer nucleonic EOSs, and the predicted median $g_1$ of a star with $\FD=0.1$ ($\mx=1$~GeV) lies $\qfRise\%$ above that of a nucleonic star, of which the direct shift accounts for about $\qfDirFrac\%$. The $68\%$ intervals of nucleonic and admixed stars overlap almost completely: the frequency of an admixed star is set mainly by the EOS that the star requires.

\begin{figure}[t]
	\includegraphics[width=\columnwidth]{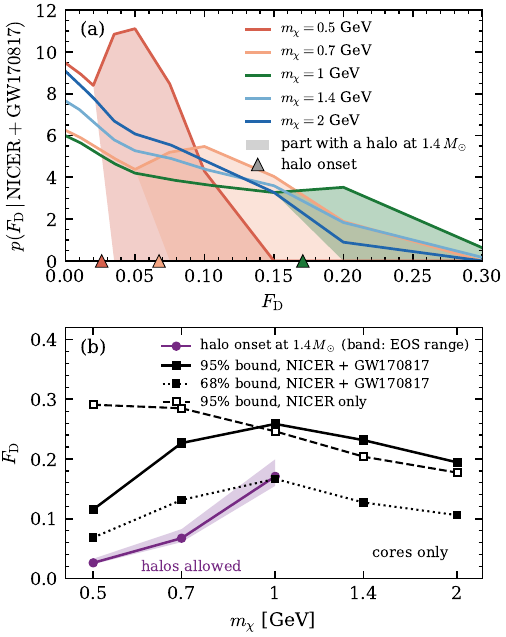}
	\caption{(a) Posterior of the dark mass fraction from NICER and GW170817 for five dark-particle masses. Shaded: the part of the posterior in which the $1.4\,\Msun$ star has a dark halo; triangles: posterior-weighted $\FD$ at which halos appear. (b) Halo onset at $1.4\,\Msun$ (line; band: range over the EOSs) and upper bounds on $\FD$ with and without GW170817, against $\mx$. Halos survive where the onset lies below the bound.}
	\label{fig:bounds}
\end{figure}

\begin{table}[t]
\caption{Bayesian results for each dark-particle mass $m_\chi$ (prior: $38$ Skyrme and $25$ RMF EOSs with equal weight; $\sigma_{\rm D}/m_\chi=1$~cm$^2$\,g$^{-1}$). Upper bounds on $\FD$ at $68$, $90$ and $95\%$ credibility from NICER and GW170817, and at $95\%$ from NICER alone; Bayes factor of free $\FD\in[0,0.3]$ against $\FD=0$ with and without GW170817; posterior probability that a $1.4\,\Msun$ star has a dark halo; posterior-weighted $\FD$ at which the $1.4\,\Msun$ star becomes a halo.}
\label{tab:bayes}
\setlength{\tabcolsep}{0.8pt}
\begin{ruledtabular}
\begin{tabular}{ccccccccc}
$m_\chi$ & \multicolumn{3}{c}{$\FD$ bound (N+GW)} & N & \multicolumn{2}{c}{$\ln\mathcal B$} & $P_{\rm halo}$ & $\FD^{\rm halo}$\\
{[GeV]} & $68\%$ & $90\%$ & $95\%$ & $95\%$ & N+GW & N & & \\
\hline
$0.5$ & $0.068$ & $0.101$ & $0.115$ & $0.291$ & $-1.04$ & $+0.11$ & $0.76$ & $0.026$\\
$0.7$ & $0.131$ & $0.197$ & $0.227$ & $0.285$ & $-0.63$ & $-0.50$ & $0.68$ & $0.067$\\
$1$ & $0.167$ & $0.235$ & $0.259$ & $0.246$ & $-0.59$ & $-0.91$ & $0.30$ & $0.171$\\
$1.4$ & $0.127$ & $0.200$ & $0.231$ & $0.204$ & $-0.83$ & $-1.11$ & $<10^{-3}$ & --\\
$2$ & $0.106$ & $0.166$ & $0.194$ & $0.177$ & $-1.00$ & $-1.23$ & $<10^{-3}$ & --\\
\end{tabular}
\end{ruledtabular}
\end{table}

\begin{figure}[t]
	\includegraphics[width=\columnwidth]{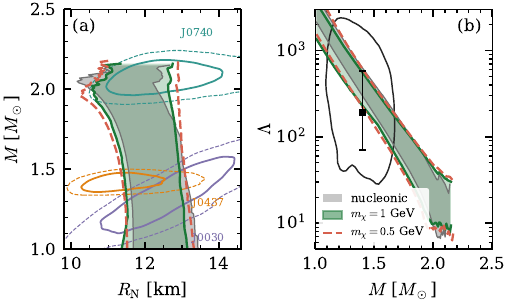}
	\caption{Posterior $90\%$ bands of (a) the mass--radius relation of the nucleonic fluid and (b) the tidal deformability, for nucleonic stars (grey), dark cores ($\mx=1$~GeV, green) and halos ($\mx=0.5$~GeV, dashed), with $\FD$ marginalized. Contours: $68/95\%$ of the NICER densities and $90\%$ of the GW170817 $(m_i,\Lambda_i)$ samples; bar: $\Lambda_{1.4}$~\cite{abbott2018prl}. Bands end where less than $20\%$ of the posterior reaches the mass.}
	\label{fig:mr}
\end{figure}

\subsection{What the data allow}

Figure~\ref{fig:bounds} and Table~\ref{tab:bayes} summarize the posterior. The data do not prefer dark matter ($\ln\mathcal B=\qplnBmin$ to $\qplnBmax$); the posterior of $\FD$ is largest at or near zero, so we quote upper bounds: $\FD<\qpUbmin$--$\qpUbmax$ at $95\%$ credibility over $\mx=0.5$--$2$~GeV. The two geometries are bounded by different data. A dark core adds mass at the centre without adding pressure to the nucleonic fluid, which is compressed: its radius and the maximum mass of the sequence fall, and the NICER radii and the $2\,\Msun$ pulsar J0740+6620 resist this. GW170817 does not, because a core also lowers the tidal deformability, which GW170817 favours. A halo acts the other way. Most of its mass lies outside the nucleonic fluid, so the nucleonic radius and the maximum mass hardly change and NICER alone barely constrains it. The tidal field, however, acts on the whole mass distribution. Just above the onset a thin halo holds too little mass at large radius to matter and lowers $\Lambda$ like a core (for SLy4 at $\mx=0.5$~GeV, $\Lambda_{1.4}$ falls from $298$ to $197$ at $\FD=0.05$), but a more massive halo extends the quadrupolar response to large radii ($\Lambda_{1.4}=359$ at $\FD=0.1$ and $1.3\times10^3$ at $0.15$), and GW170817 cuts it off. Halos form at a dark fraction that rises steeply with $\mx$, from $\FD\approx\qpOnA$ at $0.5$~GeV to $\qpOnC$ at $1$~GeV, as the Fermi scaling implies, and they survive the data only where this onset lies below the bound, for $\mx\lesssim\qpHaloMax$~GeV. The bound on cores is set by the stiffest EOSs in the prior, which can compensate the softening by a core: stiff functionals such as BigApple or BSR1--5, which GW170817 disfavours as nucleonic stars, pass it once a core softens them, so that most of the posterior at $\mx=1$~GeV ($\qpWrmf\%$) sits in stiff RMF functionals with a dark core; with the Skyrme functionals alone the $95\%$ bound would be $\qpUbSkC$ at $\mx=1$~GeV. Since all our functionals satisfy the nuclear-matter constraints at saturation, those constraints cannot tighten the bound; the pressure at a few times saturation density can. In mass, radius and tidal deformability the allowed admixed stars are indistinguishable from nucleonic ones (Fig.~\ref{fig:mr}): without dark matter the data give $R_{1.4}=\qpRlo$--$\qpRhi$~km and $\Lambda_{1.4}=\qpLlo$--$\qpLhi$ ($90\%$), and the bands of stars with cores or halos fall inside these.

\begin{figure}[t]
	\includegraphics[width=\columnwidth]{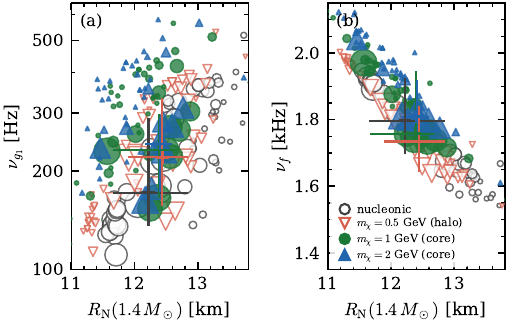}
	\caption{Posterior of (a) $\nu_{g_1}$ and (b) $\nu_f$ of $1.4\,\Msun$ stars against the radius of the nucleonic fluid. Each point is one EOS, with area proportional to its posterior weight, for nucleonic stars and for stars with $\FD=0.1$; crosses: posterior median and $68\%$ interval.}
	\label{fig:family}
\end{figure}

\subsection{Degeneracy with the symmetry energy}

The relevant comparison is therefore between stars that NICER and GW170817 cannot tell apart. Among all pairs of a nucleonic and an admixed $1.4\,\Msun$ star of the $63$ EOSs whose radii agree to $0.15$~km and whose tidal deformabilities agree to $10\%$ ($\qmTwN$ pairs), the admixed star has the higher $g_1$ frequency in $\qmTwAbove\%$, by a factor of $\qmTwMed$ at the median, and the lower one in the rest. The admixed star needs a stiffer nucleonic EOS to reach the same radius, and a stiffer EOS usually, but not always, comes with a larger slope $L$, a steeper proton fraction and stronger buoyancy. Figure~\ref{fig:family}(a) shows the consequence for the posterior: the stars with dark cores sit at higher $g_1$ than the nucleonic stars of the same radius, but within the spread of the nucleonic stars. That spread is the symmetry energy. A relation in the radius alone describes the nucleonic $g_1$ to $\qmFitRPct\%$; adding $J$ and $L$ reduces this to $\qmFitPct\%$ for all $63$ functionals, and in that relation a core with $\FD=0.1$ is equivalent to raising $L$ by $\qmdLC$--$\qmdLE$~MeV at fixed radius and $J$ ($\mx=1$--$2$~GeV), a fraction of the present uncertainty of $L$, $\approx28$~MeV~\cite{oertel2017}. The $g_1$ frequency therefore measures the density dependence of the symmetry energy more than the dark matter, which makes quantitative, for data-selected stars, the degeneracy that Ref.~\cite{shirke2025} noted for the one-fluid model.

\subsection{The $f$-mode test}

The $f$-mode is a different kind of oscillation: it is restored by pressure rather than buoyancy, and its frequency follows the mean density of the nucleonic fluid. At fixed radius a core, which compresses the nucleonic fluid, raises it [Fig.~\ref{fig:family}(b)], and at fixed EOS the core raises $\nu_f$ by $\qmDfCmin$--$\qmDfCmax\%$ ($\mx=1$~GeV) for every functional, whereas the $g_1$ shift varies from $\qmDgCmin$ to $\qmDgCmax\%$. In the posterior, however, the data again select a stiffer EOS, and the $f$-mode and radius together separate cores from nucleonic stars only at the level of a few percent.

\begin{figure}[t]
	\includegraphics[width=\columnwidth]{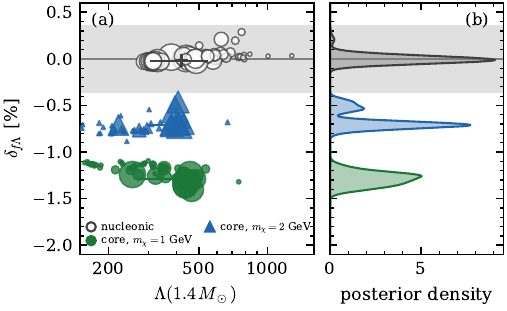}
	\caption{Offset from the $f$--Love relation of Ref.~\cite{chan2014}, $\delta_{f\Lambda}=M\omega_f/F(\ln\Lambda)-1$, with $\omega_f$ the full-GR $f$-mode of the nucleonic fluid and $\Lambda$ the tidal deformability of the whole star, for $1.4\,\Msun$ stars: (a) against $\Lambda$, points as in Fig.~\ref{fig:family}; (b) posterior distribution. Grey band: $1\sigma$ residual of the relation over $30\,720$ single-fluid EOSs, $0.37\%$~\cite{chu2026}.}
	\label{fig:flove}
\end{figure}

A second test removes the need for a radius. The relation between $M\omega_f$ and $\ln\Lambda$~\cite{chan2014} holds for single-fluid stars to $0.37\%$~\cite{chu2026}, and the posterior of our nucleonic stars follows it to within $\qmDflNuc\%$ (Fig.~\ref{fig:flove}). A dark component breaks it for a simple reason: $\Lambda$ responds to the whole mass distribution, whereas the $f$-mode lives in the nucleonic fluid. For one fluid both quantities are fixed by the same density profile, which is why the relation is nearly universal. A dark core concentrates mass at the centre, which lowers the Love number and $\Lambda$ of the whole star, while the $f$-mode rises only as much as the compressed nucleonic fluid demands; compared with a nucleonic star of the same $\Lambda$, the admixed star has the lower $f$-mode frequency. Stars with dark cores of $\FD=0.1$ fall below the relation by $\qmDflC\%$ ($\mx=1$~GeV) and $\qmDflE\%$ ($\mx=2$~GeV) at the posterior median, and none of them lies within $\qmDflEmax\%$ of it; at fixed EOS the offset from a core is nearly the same for every functional ($\qmDDflCmin$--$\qmDDflCmax\%$ for $\mx=1$~GeV). This extends the Cowling result of Ref.~\cite{sotani2025b} for dark fractions $\ge0.2$ to full GR and to the small fractions the data allow. The test compares two observables of the same star, $\Lambda$ from the inspiral and $\nu_f$ from the dynamical tide or a post-merger signal; because $M\omega_f$ depends only weakly on $\Lambda$, a $1\%$ offset corresponds to a few-percent change in $\Lambda$, a target for third-generation detectors.

\subsection{Mode data demand a complete EOS prior}

Could a measured mode frequency reveal dark matter? We added to the likelihood a mock $g_1$ frequency of a $1.4\,\Msun$ star from one of the twin pairs ($\mx=1$~GeV, $\FD\le0.1$). With a $5\%$ error, which stands for the theory systematic (the crust--core transition alone moves $\nu_{g_1}$ by several percent), the admixed star gives $\ln\mathcal B=\qfcGD$ and its nucleonic partner $\qfcGN$ at the median; with $g_1$ and $f$ measured to $1\%$ the separation becomes decisive ($\qfcGFD$ against $\qfcGFN$). This assumes that the true nucleonic EOS is in the prior. When it is left out, $\qlooGfrac\%$ of nucleonic stars give $\ln\mathcal B>1$ for dark matter with $g_1$ at $5\%$, and $\qlooGFfrac\%$ (up to $\ln\mathcal B=\qlooGFmax$) with $g_1$ and $f$ at $1\%$, because a dark core is then the model's only way to reproduce mode frequencies that no EOS of the prior has. Percent-level mode data can manufacture evidence for dark matter as easily as they find it.

\subsection{Detectability}

In a binary, the resonant excitation of $g_1$ shifts the gravitational-wave phase~\cite{lai1994,ho1999,gittins2026}. For two $1.2\,\Msun$ stars the posterior gives a median of $\qdphiNuc$~rad without and $\qdphiCore$~rad with dark cores ($\mx=1$~GeV, $\FD=0.1$), and $95$th percentiles of $\qdphiNuchi$ and $\qdphiCorehi$~rad, near the $\approx0.03$~rad that the Einstein Telescope can detect in favourable events~\cite{gittins2026}. As free oscillations the $g$-modes are hard to see: gravitational radiation damps them only over $10^{\qTauLo}$--$10^{\qTauHi}$~s ($90\%$ posterior interval for nucleonic stars), and the shear and bulk viscosity of normal matter~\cite{lai1994,andersson2001}, evaluated on the full-GR eigenfunctions, damp them faster at every core temperature we considered. Our frequencies apply above the neutron critical temperature; in superfluid cores the composition $g$-modes are carried by the muons or disappear~\cite{kantor2014}.

\section{Limitations}

Several approximations limit the numbers, though we expect none of them to change the conclusions. (i) \textit{Crust.} All our crusts are unstratified ($\cad=\ceq$). The crust--core transition alone moves the $g_1$ frequency: for SLy4 at $1.4\,\Msun$ the Cowling $\nu_{g_1}$ is $128.7$, $132.6$ and $139.8$~Hz for a transition at $n_t=0.089$, $0.070$ and $0.051$~fm$^{-3}$. This systematic, comparable to the direct dark-matter shift, is why we took $5\%$ as the theory error of $\nu_{g_1}$; a stratified crust treated consistently with the core~\cite{gao2025} would reduce it. (ii) \textit{Superfluidity.} Our modes are those of normal $npe\mu$ matter. Below the neutron critical temperature the composition $g$-modes of $npe$ matter disappear, and with muons temperature-dependent superfluid $g$-modes of several times higher frequency appear~\cite{kantor2014} (see also Ref.~\cite{passamonti2016}). The dark matter acts on the nucleons only through $\Phi'$ and through the nucleonic EOS that the data select, neither of which involves the superfluid dynamics, so we expect the degeneracy with the symmetry energy to carry over, but we have not computed it. (iii) \textit{EOS prior.} The prior is a discrete set of $63$ functionals, so the posterior summaries are coarse, and the bound on dark cores depends on the stiffest functionals admitted, as shown above. A continuous prior built from nuclear empirical parameters~\cite{margueron2018}, with the proton fraction, and hence $D$, computed self-consistently~\cite{danarianto2026}, would remove this dependence; it must control the high-density symmetry energy that sets $\nu_{g_1}$. (iv) \textit{Pulse profiles.} The NICER analyses assume a vacuum exterior. The dark mass of a halo outside $R_{\rm N}$ bends the light~\cite{miao2022,shawqi2024,shawqi2026}, so the likelihood of halo models is approximate. (v) \textit{Dark-matter model.} We used one fermionic model with one self-interaction, a single $\FD$ for all stars, zero temperature, no rotation and coupling through gravity only. A dark fraction that varies from star to star would weaken the bounds, and other dark sectors, such as bosonic~\cite{rutherford2023} or mirror~\cite{hippert2023} dark matter, change the dark EOS but not the mechanism, because the $g$-modes live in the nucleonic fluid and feel the dark component only through $\Phi'$.

\section{Conclusions}

For dark matter coupled to neutron stars only by gravity, the composition $g$-modes are a poor dark-matter probe. The dark matter reaches the buoyancy only through the gravitational field, so its direct effect on the $g$-modes is several times below the one-fluid estimate. The larger effect comes through the nucleonic EOS that the data demand once the star carries dark matter, and because the $g$-modes are set by the symmetry energy, it is degenerate with nuclear physics: a core with $10\%$ of the mass is worth $\qmdLC$--$\qmdLE$~MeV of $L$. The $f$-mode, which follows the mean density, responds to a dark core in the same way for all functionals, and the $f$--Love relation, which nucleonic stars obey closely, is broken by about one percent, a test that uses observables of a single star. A dark-matter interpretation of any mode frequency requires a nucleonic EOS prior that is complete at the level of the measurement.

\begin{acknowledgments}
	Authors thank Ankit Kumar and Hajime Sotani for their suggestions and insights on the two-fluid model that helped shape this work.
\end{acknowledgments}

\bibliography{refs}
\end{document}